# Q Dependence of Magnetic Resonance Mode on FeTe$_{0.5}$Se$_{0.5}$ Studied by Inelastic Neutron Scattering

**Motoyuki Ishikado [1],\*, Katsuaki Kodama [2], Ryoichi Kajimoto [3], Mitsutaka Nakamura [3], Yasuhiro Inamura [3], Kazuhiko Ikeuchi [1], Sungdae Ji [1],† and Masatoshi Arai [3],‡ and Shin-ichi Shamoto [4],\***

[1] Neutron Science and Technology Center, Comprehensive Research Organization for Science and Society (CROSS), Tokai, Ibaraki 319-1106, Japan
[2] Materials Sciences Research Center, Japan Atomic Energy Agency (JAEA), Tokai, Ibaraki 319-1195, Japan
[3] J-PARC Center, Japan Atomic Energy Agency (JAEA), Tokai, Ibaraki 319-1195, Japan
[4] Advanced Science Research Center, Japan Atomic Energy Agency (JAEA), Tokai, Ibaraki 319-1195, Japan
\* Correspondence: m_ishikado@cross.or.jp (M.I.); shamoto.shinichi@jaea.go.jp (S.-i.S.)
† Current address: Max Planck POSTECH Center, 37673 Pohang, Korea.
‡ Current address: European Spallation Source ESS AB, Stora Algatan 4, SE-22100 Lund, Sweden.

**Abstract:** Inelastic neutron scattering measurements have been performed on a superconducting single crystal FeTe$_{0.5}$Se$_{0.5}$ to examine the **Q**-dependent enhancement of the dynamical structure factor, $S(\mathbf{Q}, E)$, from $\mathbf{Q} = (0, 0)$ to $(\pi, \pi)$, including $(\pi, 0)$ in the superconducting state. In most of iron-based superconductors, $S(\mathbf{Q}, E)$ is enhanced at $\mathbf{Q} = (\pi, 0)$, where the "magnetic resonance mode" is commonly observed in the unfolded Brillouin zone. Constant-$E$ cuts of $S(\mathbf{Q}, E)$ suggest that the enhancement is not uniform in the magnetic excitation, and limited around $\mathbf{Q} = (\pi, 0)$. This result is consistent with the theoretical simulation of the magnetic resonance mode due to the Bardeen–Cooper–Schrieffer coherence factor with the sign-reversing order parameter of s$_{\pm}$ wave.

## 1. Introduction

After the discovery [1], iron-based superconductors have attracted a significant amount of attention as new unconventional superconductors subsequent to cuprates, heavy fermions, organic condutors, etc. The following are the characteristics of iron-based superconductors: (1) second-highest superconducting transition temperature ($T_c$), next to cuprates, (2) superconductivity occurs in FePn (=pnictide) or FeCh (=chalcogenide) plane, (3) multiband system of iron 3d electrons (resulting in multi-gap superconductivity [2]), and (4) unconventional pairing mechanism. In particular, for (4), it is theoretically pointed out that the high $T_c$ of the iron-based superconductors cannot simply be explained by conventional electron-phonon coupling [3] in the early stage. Instead, the spin fluctuation, orbital fluctuation, and nematic scenarios are under dispute as possible pairing mechanisms. Depending on the scenario, the symmetry of the superconducting order parameter becomes s$_{\pm}$- or s$_{++}$-wave states [4–7].

Fermi surfaces (FSs) of iron-based superconductors consist of hole FSs located at $\mathbf{k} = (0, 0)$ (so called Γ point) and electron FSs located at $\mathbf{k} = (\pi, 0)$ (so called M point) in the unfolded Brillouin zone (one Fe in the unit cell), and the nesting between the hole and the electron FSs leads to the spin fluctuation. Then, the superconductivity occurs from electron pairing mediated by the spin fluctuation. In this case, order parameter changes its sign between Γ and M point FSs, resulting in no nodes. Hence, this is called an $s_\pm$ wave, which was theoretically proposed in the early stage of the study [4,5]. On the other hand, the pairing state called $s_{++}$, in which the order parameters at Γ and M points have the same sign, was later proposed [8]. This paring state is formed by the fluctuation of the electron occupation in Fe 3d orbitals (so-called orbital fluctuation). Note that in multi-gap superconductivity, the pairing state is determined not only by the intra-band spin or phonon-mediated pairing, but also by the exchange interaction between the multiple condensates [9,10]. Experimentally, for example, Scanning Tunneling Microscopy (STM)/Scanning Tunneling Spectroscopy (STS) measurements have detected the quasi-particle interference generated by $s_\pm$ wave pairing [11]. On the other hands, there are reports that $T_c$ suppression by the impurity is consistent with $s_{++}$ [12]. Symmetry transition from $s_\pm$ to $s_{++}$ is also observed in $K_xFe_{2-y}(Se_{1-z}S_z)_2$ system [13]. In this way, there seem to exist several reports to support both pairing symmetry.

The enhancement of the generalized dynamical spin susceptibility, $\chi''(E)$, at $\mathbf{Q} = (\pi,0)$, in the superconducting state, has been observed by the inelastic neutron scattering (INS) measurements for various iron-based superconductors [14–18]. The enhancement is called the "magnetic resonance mode," and it was considered to be the evidence for $s_\pm$-wave. In the superconducting state, the quasi-particle density near the energy gap Δ is enhanced, and the $\chi''(E)$ includes BCS (Bardeen–Cooper–Schrieffer) coherence factor. Because the signs of the order parameters at Γ and M points are opposite in $s_\pm$-wave superconductors, the coherence factor becomes 1 near the gap energy, resulting in the enhancement of $\chi''(E)$ at $\mathbf{Q} = (\pi, 0)$ and $E \sim 2\Delta$. The resonance enhancement of $\chi''(E)$ observed at $\mathbf{Q} = (\pi, 0)$ and $E \sim 2\Delta$, for various iron-based superconductors, can be explained by the above scenario.

However, later, Onari and his collaborators showed that the enhancement of $\chi''(E)$ can be reproduced even in the $s_{++}$-wave superconductor by taking account of the large quasi-particle damping in the normal state [8,19]. In their calculation, the $\chi''(E)$, at $\mathbf{Q} = (\pi, 0)$ and at $E \sim 2\Delta$, was much smaller than the $\chi''(E)$ calculated in the $s_\pm$ wave superconductor because the coherence factor becomes zero near the gap energy in $s_{++}$ wave superconductor. However, the $\chi''(E)$ in the normal state is suppressed by the quasi-particle damping relative to the $\chi''(E)$ in the superconducting state. As a result, the enhancement of $\chi''(E)$ at $\mathbf{Q} = (\pi, 0)$, below $T_c$, can also be reproduced in the $s_{++}$ scenario. Moreover, according to their result, the differences between $s_\pm$ and $s_{++}$ are the peak energy and the width. The broad enhancement is expected to occur above 2Δ for $s_{++}$-wave due to the dissipationless quasi-particle damping effect [8], and the sharp resonant peak appears below 2Δ by the opening the gap for $s_\pm$-wave states [20]. Width of the peak depends on the quasi-particle damping parameter. For example, the quasi-particle damping parameter of Fe(Te,Se) can be estimated to be about 2–3 meV at the superconducting state based on an angle resolved photoemission spectroscopy measurement [21]. The width depends on a material or doping concentration as shown in sulfur doped FeSe [22]. In addition, the superconducting gap value depends on the experimental methods, resulting in the ambiguity of the estimation of the peak energy. Therefore, it is difficult to distinguish between $s_\pm$ and $s_{++}$ experimentally by measuring $\mathbf{Q} = (\pi,0)$ INS data.

The **Q**-dependent enhancements of the $\chi''(\mathbf{Q}, E)$ in the $s_\pm$ and $s_{++}$-wave superconducting states are theoretically calculated by Nagai and Kuroki [23,24]. According to their simulation, the enhancement appears only around $(\pi, 0)$ for $s_\pm$-wave due to the sign-reversing order parameter. On the other hand, the enhancement appears uniformly at any **Q** position for $s_{++}$-wave by the quasi-particle damping effect.

This theoretical simulation enables us to distinguish whether the superconducting symmetry is $s_\pm$ wave or $s_{++}$ wave, by measuring the dynamical structure factor, $S(\mathbf{Q}, E)$, in the superconducting and non-superconducting states. We performed INS measurements on a $FeTe_{0.5}Se_{0.5}$ single crystal in order to study the **Q** dependence of the $S(\mathbf{Q}, E)$ spectra from $\mathbf{Q} = (0,0)$ to $(\pi,\pi)$, including $(\pi,0)$. Good statistics are required to measure the **Q**-dependent enhancement and to study in detail with new approach and method. In this sense, Fe(Te,Se) system is suitable as a target superconductor, because large single crystals can be grown although its magnetic excitations have been well studied and established so far [25,26].

## 2. Experimental Results

Figure 1a,b are two-dimensional dynamical structure factor, $S(\mathbf{Q}, E)$, for the $(H, K)$ plane in the normal state (a) at $T = 20$ K and superconducting state (b) at $T = 5$ K. The $S(\mathbf{Q}, E)$ maps are integrated in the energy ($E$) range of $7 \leq E \leq 10$ meV. In the normal state, in Figure 1a, we can see finite intensity at four-fold symmetric $\mathbf{Q}$ positions of $(\pi, 0)$. In the superconducting state, these signals are enhanced as shown in Figure 1b.

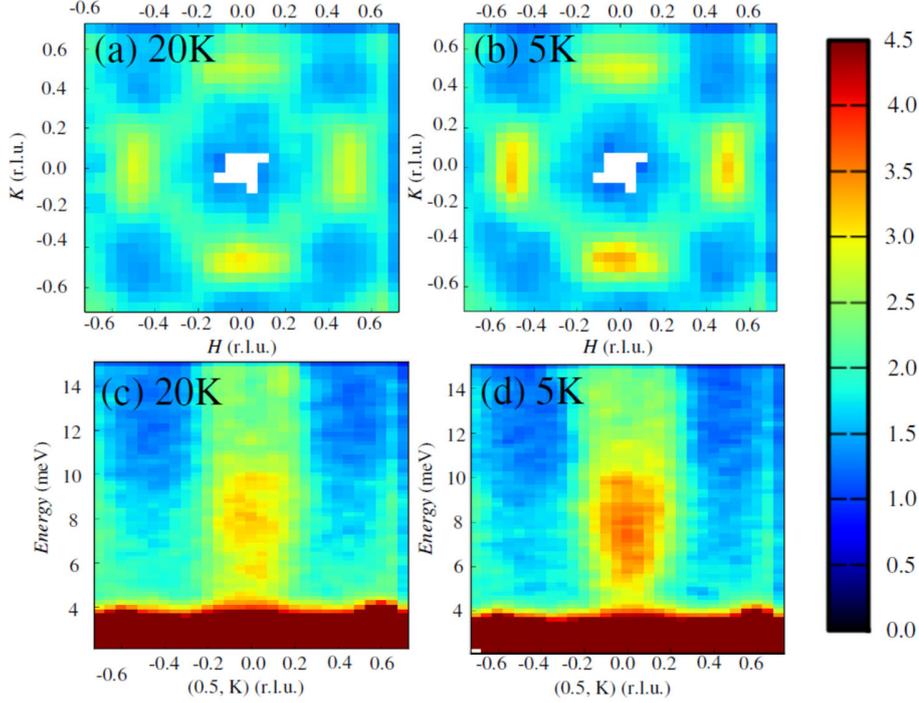

**Figure 1.** Dynamical structure factor $S(\mathbf{Q}, E)$ of FeTe$_{0.5}$Se$_{0.5}$ for $(H, K)$ plane at (**a**) $T = 20$ K and (**b**) $T = 5$ K with $E_i$ = 45.1 meV, and the integrated energy range is $7 \leq E \leq 10$ meV. $S(\mathbf{Q}, E)$ for $(0.5, K)$ at (**c**) $T = 20$ K and (**d**) $T = 5$ K. Intensity scale given by the color bar is same for all figures (**a–d**).

The $E$ dependence of $S(\mathbf{Q}, E)$ along $(0.5, K)$ is shown in Figure 1c,d. The cut width was $\Delta H = \pm 0.15$ (r.l.u.). The data of the four-fold symmetric directions were averaged. In the normal state spectra, Figure 1c, finite magnetic scattering is observed around $(\pi, 0)$, and its intensity is particularly high around 8 meV. In the superconducting state at 5 K in Figure 1d, this intensity is clearly enhanced. To see further details at $(\pi, 0)$, we performed constant-$Q_K$ cuts at $(\pi, 0)$ with a width of $\Delta K = \pm 0.05$ (r.l.u.), which are shown in Figure 2a. Constant-$Q_K$ cuts at $(\pi, \pi)$ are also shown in Figure 2b. The enhancement around 8 meV is observed at $(\pi, 0)$ in the superconducting state in Figure 2a, whereas the intensity is slightly suppressed at $E \sim 6$ meV at $(\pi, \pi)$ in Figure 2b.

To compare the $\mathbf{Q}$ dependence of the observed $S(\mathbf{Q}, E)$ with the theoretically simulated $\mathbf{Q}$ dependence [24], we performed constant-$E$ cuts along $\mathbf{Q} = (0,0)$ to $(\pi, 0)$, and to $(\pi, \pi)$ at $E = 8$ meV where the $S(\mathbf{Q}, E)$ is largely enhanced in the superconducting state. The cut widths were $\Delta E = \pm 1.6$ meV and $\Delta H, \Delta K = \pm 0.05$ (r.l.u.). The data were averaged with four-fold symmetric cuts. To estimate the background, we fitted the $\mathbf{Q}$-scan cuts along $(H, 0)$, and $(0, K)$ or other symmetric cuts in a $\mathbf{Q}$ range from 0.15 to 1.5−2.0 (r.l.u.) at both 5 K and 20 K by a combination of constant, $\mathbf{Q}^2(=Q_H^2 + Q_K^2 + Q_L^2)$, and Gaussian functions. They are attributed to constant background, phonons, and magnetic scattering, respectively. The constant background is subtracted from the constant-$E$ cut intensity for both 5 and 20 K. The extracted scattering intensity at $E$ = 8 meV is shown in Figure 2c. The $\mathbf{Q}^2$ term in the background is also shown in Figure 2c. We can see that the intensity at $(\pi, \pi)$ is very weak. It is mostly attributed to the $\mathbf{Q}^2$ background which decreases as $\mathbf{Q}$ moves away from $(\pi, \pi)$ as shown in Figure 2c. Here, the magnetic resonance mode is observed only around $\mathbf{Q} = (\pi,0)$ in the superconducting state. It disappears as $\mathbf{Q}$ departs from $\mathbf{Q} = (\pi,0)$.

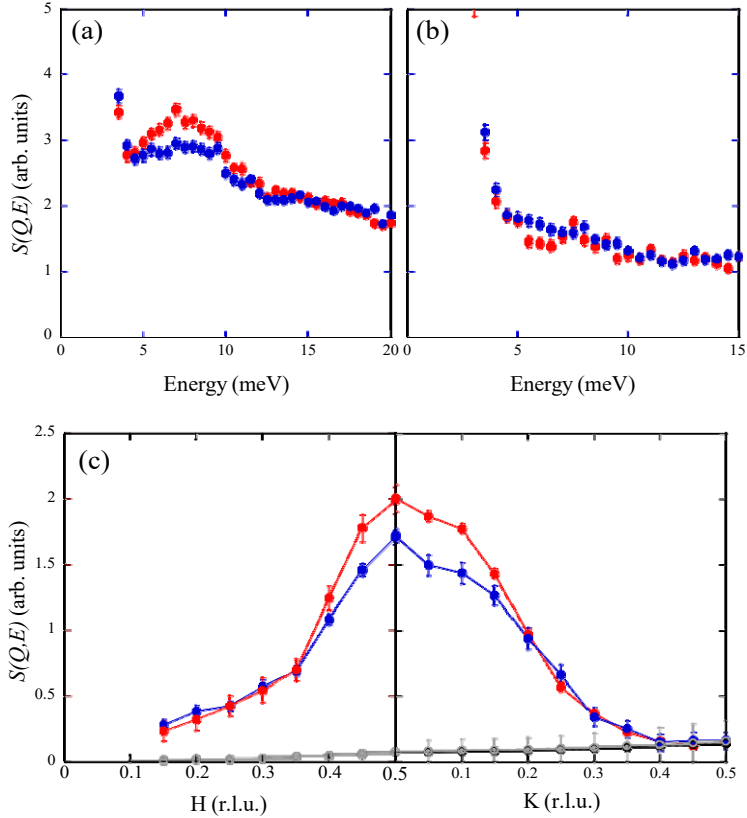

**Figure 2.** Constant-$\mathbf{Q}$ cuts of $S(\mathbf{Q}, E)$ at (**a**) $\mathbf{Q} = (\pi, 0)$ and (**b**) $\mathbf{Q} = (\pi, \pi)$. Cut width is $\Delta H$, $\Delta K = \pm 0.05$ (r.l.u.). Red and blue closed circles are measured at $T$ = 5 K and 20 K, respectively. (**c**) Line cuts of $S(Q, E = 8\text{meV})$ along $\mathbf{Q} = (0,0)$ to $(\pi,0)$ and to $(\pi,\pi)$. Red and blue closed circles are measured at $T$ = 5 K and 20 K, respectively. $\mathbf{Q}^2$ backgrounds are also shown with black (5 K) and grey (20 K) open circles, respectively.

## 3. Discussion

In the case of the $s_\pm$ wave, the sign of the order parameter is reversed between the hole ($\Gamma$ point) and the electron (M point) FSs. On the contrary, there is no sign change between them in the $s_{++}$ case. In this configuration of the FSs, two scattering processes can be considered. One is the scattering vector $\mathbf{Q} = (\pi, 0)$ bridging $\Gamma$ and M FSs, where the "magnetic spin resonance" is commonly observed in iron-based superconductors. Figure 2a shows the enhancement at the corresponding scattering process. The other scattering process that bridges the electron-electron FSs is around $\mathbf{Q} = (\pi, \pi)$. The resonance mode with ~ $\mathbf{Q} = (\pi, \pi)$, corresponding to the latter process, was observed in $Rb_xFe_{2-y}Se_2$ superconductor without the $\Gamma$ Fermi surface [27]. Figure 2b corresponds to the scattering process between M-M point FSs, although its intensity is very weak.

Before examining the $\mathbf{Q}$ dependence of the magnetic resonance mode, let us reconsider the theoretical mechanism. In the following, the dynamical spin susceptibility $\chi''(Q, E)$ is used instead of the dynamical structure factor $S(Q, E)$, which is simply converted by using the Bose factor. However, the Bose factor does not affect the mechanism because its effect is negligible. As described briefly in the introduction, the BCS coherence factor is included in the dynamical spin susceptibility of the superconducting state:

$$\frac{1}{2}\left(1-\frac{\epsilon_{\mathbf{k}}\epsilon_{\mathbf{k+q}}+\Delta_{\mathbf{k}}\Delta_{\mathbf{k+q}}}{E_{\mathbf{k}}E_{\mathbf{k+q}}}\right) \approx \frac{1}{2}\left(1-\frac{\Delta_{\mathbf{k}}\Delta_{\mathbf{k+q}}}{|\Delta_{\mathbf{k}}||\Delta_{\mathbf{k+q}}|}\right), \qquad (1)$$

where $E_{\mathbf{k}} = \sqrt{\epsilon_k^2 + \Delta_k^2}$ is the quasi-particle dispersion relation and $\varepsilon_{\mathbf{k}}$ is the band energy measured relative to Fermi energy. The exact formula of the BCS coherence factor, the left-hand side of Equation (1), is approximated to the right-hand side of Equation (1) near the Fermi energy. In the superconducting state, the density of the quasi particle near $\Delta_{\mathbf{k}}$, is enhanced; the enhancement is accompanied by the opening of the gap. In the case of $\Delta_{\mathbf{k}} = -\Delta_{\mathbf{k+q}}$, the coherence factor becomes almost 1, and the $\chi''(E)$ is enhanced below $T_c$. This is observed as "magnetic resonance". In the case of $\Delta_{\mathbf{k}} = \Delta_{\mathbf{k+q}}$, $\chi''(E)$ is much smaller than the $\chi''(E)$ for $\Delta_{\mathbf{k}} = -\Delta_{\mathbf{k+q}}$, because the coherence factor becomes almost zero. However, as mentioned in the introduction, if the quasi-particle damping $\gamma$ largely suppresses the $\chi''(E)$ in the normal state, the $\chi''(E)$ can increase below $T_c$. For such a large $\gamma$, we can expect enhancement of $\chi''(E)$ for both s++ and s±-wave superconductors. However, the enhancement of $\chi''(E)$ at $\mathbf{Q} = (\pi, 0)$ for s± wave increases more than that of s++ wave by the BCS coherence factor due to the sign reversal ($\Delta_{\mathbf{k}} = -\Delta_{\mathbf{k+q}}$). On the contrary, $\chi''(E)$ at $\mathbf{Q} = (\pi, \pi)$ for s++ and s±-wave superconductors exhibit the same T-dependent enhancement, because both superconducting symmetries satisfy the conditions of $\Delta_{\mathbf{k}} = \Delta_{\mathbf{k+q}}$ at this $\mathbf{Q}$ position. For a small $\gamma$ value, which does not largely suppress the $\chi''(E)$ in the normal state, the $\chi''(E)$ at $\mathbf{Q} = (\pi, 0)$ of s++ wave decreases, whereas the $\chi''(E)$ at $\mathbf{Q} = (\pi, 0)$ of s± wave increases below $T_c$. There is no enhancement of $\chi''(E)$ at $\mathbf{Q} = (\pi, \pi)$ because the sign does not change ($\Delta_{\mathbf{k}} = \Delta_{\mathbf{k+q}}$) for both s++ and s±-wave. Consequently, the $\chi''(E)$ at $\mathbf{Q} = (\pi, \pi)$ below $T_c$ is enhanced for large $\gamma$ but suppressed for small $\gamma$. Therefore, by checking at $(\pi, \pi)$, we can distinguish whether the quasi-particle damping is small or large. The above discussion for $\mathbf{Q} = (\pi, 0)$ and $(\pi, \pi)$ can be extended to a more general discussion of $\mathbf{Q}$ dependence as shown by Nagai and Kuroki [24]. The magnitude of $\gamma$ and the superconducting symmetry can be determined by examining the $\mathbf{Q}$ dependence of the enhancement of $\chi''(E)$. Nagai and Kuroki showed that the enhancement structure localizes around $\mathbf{Q} \sim (\pi, 0)$ in the case of s± wave, while it is uniform or nearly independent of $\mathbf{Q}$ in the case of s++ wave with large damping [24]. To investigate the $\mathbf{Q}$ dependence of the enhancement, taking the superconducting-to-normal ratio of $\chi''(E)$ is useful.

Figure 3 shows the superconducting-to-normal state ratio of $S(\mathbf{Q}, E)$ along the $\mathbf{Q} = (0,0)$ to $(\pi,0)$, and to $(\pi,\pi)$. If the enhancement occurs in the superconducting state, the ratio is greater than 1. The ratio is greater than 1 around $\mathbf{Q} \sim (\pi, 0)$, however, it is nearly 1 or slightly less than 1 away from $\mathbf{Q} = (\pi, 0)$ within the experimental accuracy. This $\mathbf{Q}$ dependence is only weakly dependent on $E$ as shown in Figure 2b. Then, the experimental results suggest that the quasi-particle dampling is small, and the superconducting symmetry of FeTe$_{0.5}$Se$_{0.5}$ is considered to be s± wave.

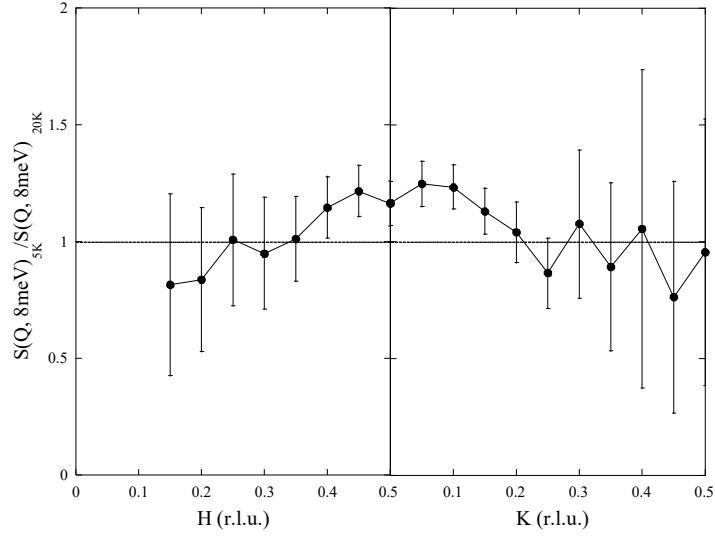

**Figure 3.** Superconducting-to-normal state ratio of the dynamical structure factor, $S(\mathbf{Q}, E)_{5K}/S(\mathbf{Q}, E)_{20K}$, at $E = 8$ meV from $\mathbf{Q} = (0, 0)$ to $(\pi, 0)$ and to $(\pi, \pi)$. The dotted line depicts a ration of 1.

Let us discuss the broadness of the resonance. Broad spin resonance is widely observed in iron-based superconductors, for example, $(Ba,K)Fe_2As_2$, $Ba(Fe,Co)_2As_2$ and $Fe(Te,Se)$ [14,28–30]. The spin resonance of $Fe(Te_{0.5}Se_{0.5})$ is also broad in $\mathbf{Q}$ and $E$ consistent with previous report [25]. Neutron scattering intensity, $S(\mathbf{Q}, E)$, observes quasi-particle excitation between hole and electron FSs, therefore it reflects multi-gap structure and quasi-particle peak width. Possibly, the broad resonance of $Fe(Te,Se)$ is caused by the broad quasi-particle peak width due to short life time and the different superconducting gap value with multiple FS structures. As an opposite example, sharp resonance is observed in $Na(Fe,Co)As$ [31]. ARPES measurements on $Na(Fe,Co)As$ reveals sharp quasi-particle peak reflecting long life time. Moreover, gap structure is almost isotropic, namely, superconducting gap value on both hole FS and electron FSs are almost the same, although the gap value itself is different in two groups [32,33]. This situation realizes sharper resonance of $Na(Fe,Co)As$ than that of $Fe(Te_{0.5}Se_{0.5})$.

Finally, we mention the resonance energy. According to the references [8,20], the enhancement appears below $2\Delta$ in $s_\pm$-wave case, and above $2\Delta$ in $s_{++}$-wave case. In the present study, the enhancement appears around 8 meV. On the contrary, $2\Delta$ of $FeTe_{0.5}Se_{0.5}$ corresponds to approximately 8 meV, according to other spectroscopic results [34,35]. Therefore, it is difficult to distinguish between $s_\pm$ and $s_{++}$-wave by only using the resonance energy. In addition, a multi-band superconductor like the present compound may have various superconducting gaps depending on the reciprocal space [2]. The early theoretical discussions assume that the Fermi surfaces at Γ and M points have the same size, but in a real case (e.g., ARPES experiments), their sizes vary depending on the materials [36–38]. The imperfect nesting and the complexity make it difficult to discuss the simple resonance energy comparison. The present analysis of $\mathbf{Q}$ dependence of magnetic resonance mode gives us a chance to determine the superconducting gap state as an alternative way.

Note here, Wang et al. observed the broad hump structure above the $2\Delta$ in the S-doped iron selenide superconductors $K_xFe_{2-y}(Se_{1-z}S_z)_2$ with z = 0.5 [13], which seems consistent with $s_{++}$ paring [13]. In this case, Nagai and Kuroki's theoretical calculation predicts the resonance enhancement over a wide $\mathbf{Q}$ range. We would like to point out that the inset of Figure 2d in Ref. [13] shows the resonance enhancement in the whole $\mathbf{Q}$ range, and this is exactly consistent with their conclusion of $s_{++}$ wave symmetry, based on our present analysis.

## 4. Materials and Methods

A single crystal of FeTe$_{0.5}$Se$_{0.5}$ was grown by the Bridgman method. Fe, Se, and Te powders were used as starting materials. The powders were ground using agate mortar in a stoichiometric ratio. Then, the powder was loaded into an alumina tube, and sealed in the evacuated quartz tube. The powders were first sintered at 600 °C for 10 h, next they were heated up to 950 °C and kept for 5 h. Then they were cooled down to 650 °C at 4 °C/h, followed by furnace cooling. The typical weight of the obtained single crystals was 10–20 g. The dc magnetic susceptibility was measured using a SQUID magnetometer (MPMS, Quantum Design Japan Inc., Tokyo, Japan) under an applied magnetic field of 10 Oe. $T_c$ was determined to be 13 K, and the superconducting shielding volume fraction reached ∼50%.

INS measurements were performed using the Fermi chopper spectrometer 4SEASONS (BL01) in J-PARC MLF (Tokai, Japan). In this measurement, we utilized the multi-$E_i$ method [39–41] with incident neutron energies of $E_i$ = 151.4, 45.1, 21.6, 12.6, and 8.2 meV, simultaneously. Note that only the data for $E_i$ = 45.1 meV are presented here. The energy resolution is 3.2 meV for $E_i$ = 45.1 meV, at $E$ = 0 meV. The total measuring time and sample weight were 25 h and 11 g, respectively at the beam power of 280 kW. The $c$-axis of FeTe$_{0.5}$Se$_{0.5}$ was set along the incident neutron beam. The measured reciprocal space is described as ($H$, $K$) because of no $L$ dependence of the intensity due to the two-dimensional magnetic excitation of FeTe$_{0.5}$Se$_{0.5}$.

## 5. Conclusions

We examined the **Q** dependence of the $S(\mathbf{Q}, E)$ by performing INS measurements on a FeTe$_{0.5}$Se$_{0.5}$ single crystal. The $S(\mathbf{Q}, E)$ is enhanced around **Q** = ($\pi$, 0) position in the superconducting state. The enhancement is gradually suppressed by going away from **Q** = ($\pi$, 0). Consequently, it is concluded that the enhancement in the superconducting state on FeTe$_{0.5}$Se$_{0.5}$ originated by multi-gap superconductivity with sign-reversing order parameter in the hole and in the electron Fermi surfaces called "s$_\pm$-wave scenario".

**Author Contributions:** Conceptualization, M.I. and S.-i.S.; formal analysis, M.I.; investigation, M.I., K.K., R.K., M.N., Y.I., K.I., S.J., M.A. and S.-i.S.; software, Y.I.; Supervision, S.-i.S.; writing—original draft preparation, M.I. and K.K.; writing—review and editing, M.I., R.K. and S.-i.S.

**Funding:** This work was supported by JST, Transformative Research-Project on Iron Pnictides (TRiP), Grant-in-Aid for Specially Promoted Research, Ministry of Education, Culture, Sports, Science and Technology (MEXT) KAKENHI Grant Number JP17001001 and Grant-in-Aid for Young Scientists (B), Japan Society for the Promotion of Science (JSPS) KAKENHI Grant Number JP15K17712.

**Acknowledgments:** The authors acknowledge Yuki Nagai, Kazuhiko Kuroki and Seiichiro Onari for fruitful discussions. The INS experiments at J-PARC MLF were carried out under Project No. 2011B0013, 2012B0130, 2012A0103, 2014A0098 and 2017A0180. The sample characterization (susceptibility measurement) was done at User Experiment Preparation Laboratory II (B402) at CROSS.

**Conflicts of Interest:** The authors declare no conflict of interest.